\documentclass[twocolumn,showpacs]{revtex4}
\usepackage{graphicx}
\usepackage{dcolumn}
\begin{document}
\title{Optical conductivity in the CuO double chains of 
PrBa$_2$Cu$_4$O$_8$:\\ Consequences of charge fluctuation}
\author{R. Amasaki$^1$}
\author{Y. Shibata$^1$}
\author{Y. Ohta$^{1,2}$}
\affiliation{$^1$Graduate School of Science and Technology, 
Chiba University, Chiba 263-8522, Japan}
\affiliation{$^2$Department of Physics, Chiba University, 
Chiba 263-8522, Japan}
\date{October 16, 2001}

\begin{abstract}
We calculate the optical conductivity of the CuO double chains 
of PrBa$_2$Cu$_4$O$_8$ by the mean-field approximation for the 
coupled two-chain Hubbard model around quarter filling.  
We show that the $\sim$40 meV peak structure, spectral shape, 
and small Drude weight observed in experiment are reproduced 
well by the present calculation provided that the stripe-type 
charge ordering presents.  
We argue that the observed anomalous optical response may be 
due to the presence of stripe-type fluctuations of charge 
carriers in the CuO double chains; the fast time scale of the 
optical measurement should enable one to detect slowly 
fluctuating order parameters as virtually a long-range order.  
\end{abstract}

\maketitle

\section{INTRODUCTION}

The CuO double chains of PrBa$_2$Cu$_4$O$_8$ (see Fig.~1(a)) 
show metallic conductivity along the chain direction down 
to 2 K \cite{horii1,horii2}.  Because the system shows 
insulating conductivity along the directions perpendicular 
to the chains (except at low temperatures $\lesssim$140 K), 
and also because the holes in the CuO$_2$ planes are localized 
in the O $2p$-orbitals around Pr ions \cite{fehrenbacher}, 
PrBa$_2$Cu$_4$O$_8$ may be regarded as a model material 
of quasi-one-dimensional (1D) correlated conductors except 
at low temperatures.  A possible Tomonaga-Luttinger-liquid 
description has then been tried to explain available 
experimental data \cite{takenaka,mizokawa}.  

A number of anomalous behaviors have however been reported 
in the charge degrees of freedom of this system 
\cite{fujiyama,takenaka}; a nuclear-quadrupole-resonance 
(NQR) experiment has demonstrated the presence of anomalous 
spin-lattice ($1/T_1$) and spin-spin ($1/T_2$) relaxations 
\cite{fujiyama} and optical conductivity experiment clearly 
shows the presence of anomalous charge excitations at 
$\omega\gtrsim 40$ meV \cite{takenaka}.  
The CuO double-chain system therefore provides a good 
opportunity for studying the anomalous charge dynamics of 
strongly correlated 1D electron systems, such as charge 
ordering (CO) and charge fluctuations, just as the quasi-1D 
organic systems do \cite{tajima,shibata}.  For example, a 
possible mechanism of metallization of this system due to 
charge frustration has recently been proposed \cite{seo}.  

Motivated by such development in the field, we in this paper 
focus on the optical conductivity spectra $\sigma(\omega)$ 
observed by Takenaka {\it et al.}~\cite{takenaka} and 
consider the origin of the appearance of the characteristic 
peak structure at $\omega\simeq 40$ meV as well as the 
presence of extremely small Drude weight.  
We use the mean-field approximation to the extended Hubbard 
model defined on a lattice of coupled two chains to obtain 
the ground-state phase diagram, where we find that there appear 
two types of CO phases, depending on the parameter values.  
We then calculate the optical conductivity spectrum
$\sigma(\omega)$ for each of these two phases in the 
mean-field approximation.  
We find that the spectra calculated for the ground state of 
the stripe-type charge ordering reproduces the experimental 
peak structure including not only its peak position and 
spectral shape but also the observed very small Drude weight.  
We also find that the spectra calculated for the in-line--type 
CO phase completely fails to reproduce the experimental 
spectra. 
%%%%%%%%%%%%%%%%%%%%%%%%%%%%%%%%%%%%%%%%%%%%%%%%%%%%%%%%%%%%%%%%%
\begin{figure}[b]
\vspace{10pt}
\begin{center}
\includegraphics[width=6.5cm,clip]{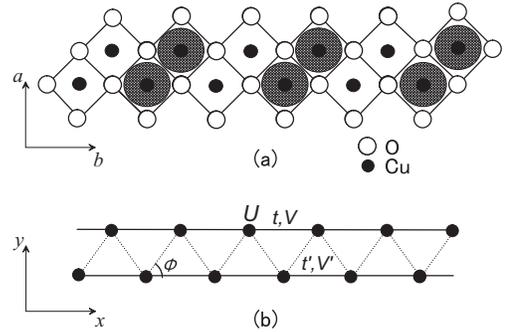}
\caption{Schematic representations of (a) the CuO double chain 
and (b) corresponding two-chain Hubbard model.  Note that the 
$x$-axis ($y$-axis) is taken along the crystallographic 
$b$-axis ($a$-axis).  The parameters $t$ and $V$ ($t'$ and 
$V'$) are associated with the chain (zigzag) bonds.  
The stripe-type CO discussed in the main text is 
schematically shown as shaded circles in (a).}  
\end{center}
\label{fig:1}
\end{figure}
%%%%%%%%%%%%%%%%%%%%%%%%%%%%%%%%%%%%%%%%%%%%%%%%%%%%%%%%%%%%%%%%%

It is known that the CuO double chains do not show the 
long-range CO but have the fluctuation of electronic charges 
with very slow time scale of the order of $0.1-10$ MHz 
\cite{fujiyama}.  
We then point out that the fast time scale of the optical 
measurement should enable one to detect slowly fluctuating 
order parameters as virtually a long-range order; i.e., the 
spectra observed should basically be the same as those for the 
long-range CO state.  Therefore, we argue that the fact that 
our mean-field optical spectra agree well with experiment 
reasonably suggest that the stripe-type charge fluctuations 
should present in the system and should be the origin of the 
$\sim$40 meV peak structure in the optical conductivity 
spectrum of the CuO double chains of PrBa$_2$Cu$_4$O$_8$.  

This paper is organized as follows.  
In Sec.~II, we define a model for describing the electronic 
states of the CuO double chains.  
In Sec.~III, we derive the mean-field phase diagram of the 
Hamiltonian, and in Sec.~IV, we calculate the optical 
conductivity of the model.  The results are discussed in 
Sec.~V by comparing with experimental data and their 
implication is considered.  Summary is given in Sec.~VI.  

\section{MODEL}

The structure of the CuO double-chain system is illustrated in 
Fig.~1(a), where we show the two CuO chains connected with 
zigzag bonds.  The electronic structure may be modeled as a 
single-band extended Hubbard-type Hamiltonian representing 
the doping holes in the antibonding band of the Cu 
$3d_{x^2-y^2}$ and O $2p_\sigma$ orbitals.  
The filling of holes in the chains is reported to be $n\sim 0.5$ 
(quarter filling) \cite{mizokawa,takenaka}; the angle-resolved 
photoemission spectroscopy experiment \cite{mizokawa} suggests 
the value $n\simeq 0.46\pm0.06$ and optical conductivity 
measurement \cite{takenaka} suggests the value $n\simeq 0.4$.  

We therefore adopt in this paper the single-band Hubbard model 
around quarter filling defined on the lattice of two chains 
connected with zigzag bonds as shown in Fig.~1(b).  
The Hamiltonian may be of the following form: 
\begin{eqnarray}
H=&-&\sum_{<ij>\sigma}t_{ij}
(c_{i\sigma}^{\dagger}c_{j\sigma}+{\rm H.c.})
+U\sum_in_{i\uparrow}n_{i\downarrow}
\nonumber \\
&+&\sum_{<ij>}V_{ij}n_in_j
\end{eqnarray}
where $c_{i\sigma}^\dagger$ ($c_{i\sigma}$) is the creation 
(annihilation) operator of a hole at site $i$ and spin $\sigma$ 
$(=\uparrow,\downarrow)$, and $n_i=n_{i\uparrow}+n_{i\downarrow}$ 
is the number operator.  We take into account the hopping parameters 
for the chain bond $t$ ($=t_2$) and for the zigzag bond $t'$ 
($=t_1$) where we note $t'\ll t$ because of the bonding angle 
minimizing the overlap of the Cu $3d_{x^2-y^2}$ and O $2p_\sigma$ 
wave functions for the hopping parameter $t'$; in the following, 
we use the value $t'/t=0.1$ confirming that the sign of $t'$ does 
not change the results.  
We also take into account the intersite Coulomb interactions $V$ 
for the chain bonds and $V'$ for the zigzag bonds, as well as 
the on-site Coulomb interaction $U$; the values are varied to 
simulate various situation.  
We restrict ourselves to the case at and less than 
quarter-filling of holes $n=N/L\le 0.5$ where $N$ and $L$ are 
the total numbers of holes and lattice sites, respectively.  

\section{MEAN-FIELD PHASE DIAGRAM}

We adopt the mean-field approximation 
\begin{eqnarray}
n_{i\uparrow}n_{i\downarrow}&\simeq&
\langle n_{i\uparrow}\rangle n_{i\downarrow}
+n_{i\uparrow}\langle n_{i\downarrow}\rangle
-\langle n_{i\uparrow}\rangle\langle n_{i\downarrow}\rangle
\\
n_in_j&\simeq&\langle n_i\rangle n_j+n_i\langle n_j\rangle
-\langle n_i\rangle\langle n_j\rangle
\end{eqnarray}
for the interaction terms of the Hamiltonian Eq.~(1).  
We assume periodicity of the ordered states where 
each unit cell $r$ contains $M$ sites; the lattice site 
may thus be specified as $i=(r,\mu)$ with the unit cell 
$r$ $(=0,\cdots,L/M-1)$ and site position $\mu$ 
$(=1,\cdots,M)$ therein.  
We have $2M$ order parameters
\begin{eqnarray}
\delta_\mu=\langle n_{\mu\uparrow}\rangle
+\langle n_{\mu\downarrow}\rangle-n
\\
2S_\mu^z=\langle n_{\mu\uparrow}\rangle
-\langle n_{\mu\downarrow}\rangle
\end{eqnarray}
with the charge neutrality constraint $\sum_\mu\delta_\mu=0$.  
We define $c_{\mu k\sigma}$ as the Fourier transform of 
$c_{i\sigma}=c_{r\mu\sigma}$: 
\begin{eqnarray}
c_{\mu k\sigma}={1\over\sqrt{L/M}}\sum_{r=1}^{L/M}
e^{ikR_{r\mu}}c_{r\mu\sigma}
\end{eqnarray}
where $R_{r\mu}=rMa+\mu a$ and $k$ is the wavevector taking 
$L/M$ values in the Brillouin zone $-\pi/Ma\le k<\pi/Ma$.  
$a$ is the lattice constant of the topologically equivalent 
single-chain model with nearest-neighbor ($t',V'$) and 
next-nearest-neighbor ($t,V$) interactions.  
The mean-field Hamiltonian is then written as
\begin{eqnarray}
H^{\rm MF}&=&\sum_{k\sigma}
\big(c_{1k\sigma}^\dagger,c_{2k\sigma}^\dagger,
\cdots,c_{Mk\sigma}^\dagger\big)
\big[H_{k\sigma}^{\rm MF}\big]
\left(
\begin{array}{ccc}
c_{1k\sigma}\\
c_{2k\sigma}\\
\vdots\\
c_{Mk\sigma}
\end{array}
\right)
\nonumber\\
&+&E_c
\end{eqnarray}
with
\begin{eqnarray}
\big[H_{k\sigma}^{\rm MF}\big]=
\big[H_t\big]+\big[H_I^{\rm MF}\big]
\end{eqnarray}
where
\begin{eqnarray}
\big[H_t\big]_{\mu\nu}=
\left\{
\begin{array}{llllllllll}
a_1^\mp & \mbox{if $\nu=\mu\pm1$}\\
a_2^\mp & \mbox{if $\nu=\mu\pm2$}\\
a_2^+   & \mbox{if $\mu=1,\nu=M-1$}\\
a_1^+   & \mbox{if $\mu=1,\nu=M$}\\
a_2^+   & \mbox{if $\mu=2,\nu=M$}\\
a_2^-   & \mbox{if $\mu=M-1,\nu=1$}\\
a_1^-   & \mbox{if $\mu=M,\nu=1$}\\
a_2^-   & \mbox{if $\mu=M,\nu=2$}\\
 0      & \mbox{otherwise}
\end{array}
\right.
\end{eqnarray}
and
\begin{eqnarray}
\big[H_I^{\rm MF}\big]_{\mu\nu}
=\Big[U\langle n_{\mu,-\sigma}\rangle
&+&V'\big(\langle n_{\mu-1}\rangle
+\langle n_{\mu+1}\rangle\big)
\nonumber\\
+V\big(\langle n_{\mu-2}\rangle
&+&\langle n_{\mu+2}\rangle\big)
\Big]\delta_{\mu\nu}.
\end{eqnarray}
In Eq.~(9), we define $a_l^\pm=-t_l\,e^{\pm ilka}$ 
$(l=1,2)$.  Also, we have
\begin{eqnarray}
E_c&=&-N\sum_{\mu=1}^M\Big[U\langle n_{\mu\uparrow}\rangle
\langle n_{\mu\downarrow}\rangle
\nonumber\\
&+&\sum_{\sigma'\sigma}\big(
V'\langle n_{\mu\sigma'}\rangle
\langle n_{\mu+1,\sigma}\rangle
+V\langle n_{\mu\sigma'}\rangle
\langle n_{\mu+2,\sigma}\rangle
\big)\Big].
\end{eqnarray}
Diagonalizing the matrix (8) by the unitary transformation
\begin{eqnarray}
\big[P_{k\sigma}\big]=\big(
{\bf p}_{k\sigma}^{(1)},
{\bf p}_{k\sigma}^{(2)},
\cdots,
{\bf p}_{k\sigma}^{(M)}
\big)
\end{eqnarray}
with ${{\bf p}_{k\sigma}^{(n)}}^\dagger
=\big({p_{1k\sigma}^{(n)}\!}^*,{p_{2k\sigma}^{(n)}\!}^*,
\cdots,{p_{Mk\sigma}^{(n)}\!\!\!}^*\big)$, we have 
\begin{eqnarray}
H^{\rm MF}=\sum_{nk\sigma}E_{k\sigma}^{(n)}
\gamma_{nk\sigma}^\dagger
\gamma_{nk\sigma}+E_c
\end{eqnarray}
where $E_{k\sigma}^{(n)}$ is the energy of the $n$-th 
quasiparticle.  The annihilation operator of the 
quasiparticle may be written as
\begin{eqnarray}
\gamma_{nk\sigma}&=&\sum_{\mu=1}^M
p_{\mu k\sigma}^{(n)*}c_{\mu k\sigma}
\\
c_{\mu k\sigma}&=&\sum_{n=1}^M{p_{\mu k\sigma}^{(n)}}
\gamma_{nk\sigma}.
\end{eqnarray}
The mean-field ground state is then written as 
\begin{eqnarray}
|0\rangle=\prod_{nk\sigma}^{\rm occ.}
\gamma_{nk\sigma}^\dagger
|{\rm vacuum}\rangle.
\end{eqnarray}
The order parameters are determined self-consistently 
so as to minimize the total energy of the system.  
The relation 
\begin{equation}
\langle n_{\mu\sigma}\rangle
={M\over L}\sum_{nk}\big|p_{\mu k\sigma}^{(n)}\big|^2
\,\theta(\varepsilon_{\rm F}-E_{k\sigma}^{(n)})
\end{equation}
is used to make iterations for self-consistency where $\theta(x)$ 
is the Fermi function at $T=0$ K and $\varepsilon_{\rm F}$ 
is the Fermi level.  

%%%%%%%%%%%%%%%%%%%%%%%%%%%%%%%%%%%%%%%%%%%%%%%%%%%%%%%%%%%%%%%%%
\begin{figure}[t]
\vspace{10pt}
\begin{center}
\includegraphics[width=7.0cm,clip]{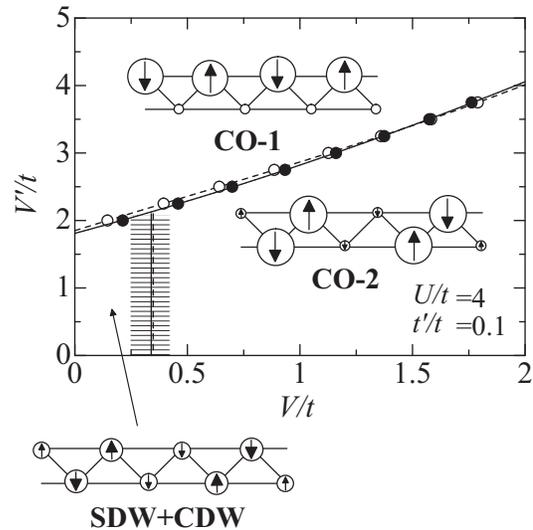}
\caption{Mean-field phase diagram of the two-chain Hubbard 
model at quarter filling ($n=0.5$, solid line with filled 
circles) and at a 10\% doping level ($n=0.45$, dashed line 
with open circles).  Two types of the charge ordering CO-1 
(in-line--type) and CO-2 (stripe-type) are illustrated 
schematically in the figure, where the circles with an arrow 
indicate the amount of charge carriers localized on the sites 
and arrows indicate the direction of the spins of carriers.  
When $t'$ is smaller than a critical value, the pure SDW 
phase appears around $V=V'=0$.  This region changes into the 
SDW+CDW (weak CO-2) phase when $t'$ becomes larger than the 
critical value (as shown in the figure).  
This phase is continuous to the CO-2 phase; 
the hatched region indicates the rapid crossover from 
the SDW+CDW phase (for small $V$) to the CO-2 phase 
(for large $V$).}
\end{center}
\label{fig:2}
\end{figure}
%%%%%%%%%%%%%%%%%%%%%%%%%%%%%%%%%%%%%%%%%%%%%%%%%%%%%%%%%%%%%%%%%
The calculated ground-state phase diagram of the model is 
shown in Fig.~2, where we use $M=8$.  
We find that there appear the following three types of phases: 
(i) The charge-ordered phase denoted as CO-1 where 
the in-line--type ordering of charges emerges (see inset of 
Fig.~2).  The spins are aligned antiferromagnetically at 
sites on one of the two chains; there is no spin polarization 
on the other chain (and even charges vanish if we set $t'=0$).  
(ii) The charge-ordered phase denoted as CO-2 where the 
stripe-type ordering of charges emerges (see inset of 
Fig.~2).  The up and down spins on the neighboring two sites 
would form a spin-singlet pair if quantum fluctuation is 
introduced.  
(iii) The spin-density-wave (SDW) phase with a small modulation 
of the charge-density distribution, which we denote as SDW+CDW, 
as shown in Fig.~2.  

We first note that the $U/t$ dependence of the phase boundary 
is rather small, that the CO phases are maintained up to 
$\sim$10\% doping levels, and that the doping dependence of 
the location of the phase boundary is also very small.  
Because we assume the unit cell of 8 sites with the 
commensurate CO, the system becomes metallic upon doping 
where the small Fermi surface emerges.  
We also note that the phase boundary between CO-1 and CO-2 
is discontinuous; i.e., the transition is of the first order.  
The transition between the phases CO-2 and SDW+CDW is 
however continuous; when $t'$ is smaller than a small critical 
value, a pure SDW phase is realized in this `SDW+CDW' 
region and the transition to the CO-2 phase is of the second 
order.  When $t'$ exceeds the critical value, the phase 
boundary becomes obscure and the transition changes into 
a crossover; when the value of $U/t$ decreases, the crossover 
becomes broader.  

The spin degrees of freedom of the fluctuating CO phases 
seems intriguing; the CO-1 phase, if it is a long-range 
CO, should behave like a 1D Heisenberg antiferromagnet, 
and in the CO-2 phase, the spin-singlet formation should 
occur.  
Experimentally, it has been found \cite{horii2} that the 
temperature-independent Pauli paramagnetic contribution 
and Curie-Weiss contribution coexist with a small anomaly 
due to the ordering of Pr spins at low temperatures.  
It would be interesting if any influence of the fluctuating 
charges on the spin states can be detected in experiment.  

\section{OPTICAL CONDUCTIVITY}

We now calculate the optical conductivity spectra for 
the states derived in the previous section.  
Using the eigenstates $|f\rangle$ and energies $E_f$ 
(where $f=0$ denotes the ground state) of the many-body 
Hamiltonian Eq.~(1), we generally define the optical 
conductivity 
$\sigma_{\alpha\alpha}(\omega)$ ($\alpha=x,y,z$) as 
\begin{eqnarray}
\sigma_{\alpha\alpha}(\omega)&=&D_{\alpha\alpha}
\delta(\omega)
\nonumber\\
&+&{{\pi e^2}\over L}\sum_{f\ne 0}{1\over\omega}
\big|\langle f|j_\alpha|0\rangle\big|^2
\delta(\omega-(E_f-E_0))
\end{eqnarray}
with the current operator
\begin{equation}
j_\alpha=i\sum_{ij\sigma}t_{ij}({\bf r}_i-{\bf r}_j)_\alpha 
c_{i\sigma}^\dagger c_{j\sigma}
\end{equation}
where $D_{\alpha\alpha}$ is the Drude weight.  
In 2D, the sum rule \cite{psaltakis,bari}
\begin{eqnarray}
\int_0^\infty\!{\rm d}\omega\,
\big(\sigma_{xx}(\omega)+\sigma_{yy}(\omega)\big)
={{\pi e^2}\over{2L}}\langle -T\rangle
\end{eqnarray} 
holds with the kinetic energy operator 
\begin{eqnarray}
T=-\sum_{ij\sigma}t_{ij}|{\bf r}_i-{\bf r}_j|^2
c_{i\sigma}^\dagger c_{j\sigma}.
\end{eqnarray}
The relation (20) may be used to calculate the Drude weight 
$D$ as 
\begin{eqnarray}
{D\over{\pi e^2}}&=&-{1\over L}\langle 0|T|0\rangle
\nonumber\\
&+&{2\over L}\sum_{f\ne 0}
{{\big|\langle f|j_x|0\rangle\big|^2}\over{E_f-E_0}}
+{2\over L}\sum_{f\ne 0}
{{\big|\langle f|j_y|0\rangle\big|^2}\over{E_f-E_0}}.
\end{eqnarray}
Note that because the present system is finite along the 
$y$-direction the Drude contribution vanishes for the electric 
field perpendicular to the chains; $D$ is thus associated 
only with $E\parallel x$.  

We use the mean-field results to calculate the optical 
conductivity and Drude weight.  The interband optical 
transition from the mean-field ground state may be written
\begin{eqnarray}
|s\leftarrow t\rangle_{k\sigma}=
\gamma_{sk\sigma}^\dagger
\gamma_{tk\sigma}|0\rangle
\end{eqnarray}
where the momentum is conserved.  We may then rewrite the 
terms in Eq.~(22) as
\begin{eqnarray}
\sum_{f\ne 0}
{{\big|\langle f|j_x|0\rangle\big|^2}\over{E_f-E_0}}
&=&\sum_{s\ne t}\sum_{k\sigma}
{{\big|_{k\sigma}\!\langle s\leftarrow t|j_x|0\rangle\,\big|^2}
\over{E_{k\sigma}^{(s)}-E_{k\sigma}^{(t)}}}
\nonumber\\
\sum_{f\ne 0}
{{\big|\langle f|j_y|0\rangle\big|^2}\over{E_f-E_0}}
&=&\sum_{s\ne t}\sum_{k\sigma}
{{\big|_{k\sigma}\!\langle s\leftarrow t|j_y|0\rangle\,\big|^2}
\over{E_{k\sigma}^{(s)}-E_{k\sigma}^{(t)}}}
\end{eqnarray}
with 
\begin{eqnarray}
&\,&_{k\sigma}\!\langle s\leftarrow t|j_x|0\rangle=
ia\cos\phi\,\big(J_{st,k\sigma}^{(1x)}+J_{st,k\sigma}^{(2x)}\big)\,
\theta_{k\sigma}^{(t)}(1-\theta_{k\sigma}^{(s)})
\nonumber\\
&\,&_{k\sigma}\!\langle s\leftarrow t|j_y|0\rangle=
ia\sin\phi\,\,J_{st,k\sigma}^{(1y)}\,
\theta_{k\sigma}^{(t)}(1-\theta_{k\sigma}^{(s)})
\end{eqnarray}
where
\begin{eqnarray}
&&J_{st,k\sigma}^{(lx)}=\sum_{\mu=1}^M
\big(
F_l\,p_{\mu+l,k\sigma}^{(s)*}\,p_{\mu k\sigma}^{(t)}-
F_l^*\,p_{\mu k\sigma}^{(s)*}\,p_{\mu+l,k\sigma}^{(t)}
\big)
\nonumber\\
&&J_{st,k\sigma}^{(1y)}=\sum_{\mu=1}^M
(-1)^\mu \big(
F_1\,p_{\mu+1,k\sigma}^{(s)*}\,p_{\mu k\sigma}^{(t)}-
F_1^*\,p_{\mu k\sigma}^{(s)*}\,p_{\mu+1,k\sigma}^{(t)}
\big)
\end{eqnarray}
and
\begin{eqnarray}
&&\theta_{k\sigma}^{(t)}=
\theta(\varepsilon_{\rm F}-E_{k\sigma}^{(t)})
\nonumber\\
&&F_l=l\,t_l\,e^{ikal}~~~(l=1, 2).
\end{eqnarray}
Here, $\phi$ is the angle between the chain and zigzag 
bonds defined in Fig.~1.  
Also, the kinetic energy term may be written 
\begin{eqnarray}
\langle 0|T|0\rangle=-a^2\sum_{nk\sigma}
\big(T_{nk\sigma}^{(1)}+2\cos^2\!\!\phi\,\,
T_{nk\sigma}^{(2)}\big)
\,\theta_{k\sigma}^{(n)}
\end{eqnarray}
with
\begin{eqnarray}
T_{nk\sigma}^{(l)}=\sum_{\mu=1}^M
\big(
F_l\,p_{\mu+l,k\sigma}^{(n)*}\,p_{\mu k\sigma}^{(n)}+
F_l^*\,p_{\mu k\sigma}^{(n)*}\,p_{\mu+l,k\sigma}^{(n)}
\big).
\end{eqnarray}
Note that in the paramagnetic state the spectral weight 
at $\omega>0$ completely vanishes, leaving only the Drude 
weight, because the present system has only a single band 
and therefore has no interband transitions.  

%%%%%%%%%%%%%%%%%%%%%%%%%%%%%%%%%%%%%%%%%%%%%%%%%%%%%%%%%%%%%%%%%
\begin{figure}[t]
\vspace{10pt}
\begin{center}
\includegraphics[width=8cm,clip]{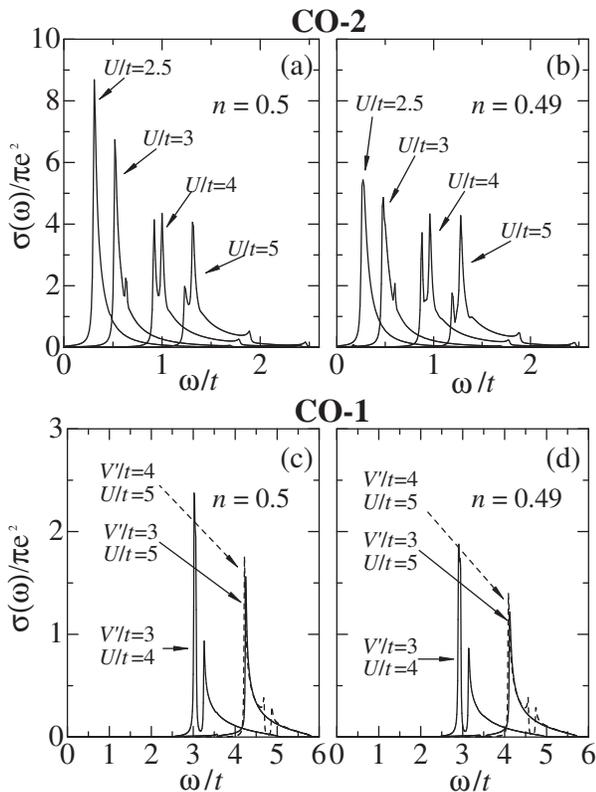}
\caption{Optical conductivity spectra $\sigma(\omega)$ with 
$E\parallel x$ calculated for the ground states of CO-2 (in 
(a) and (b)) and CO-1 (in (c) and (d)).  Drude peak 
($\omega=0$) is not included.  
The results are at quarter filling ($n=0.5$) in (a) and (c), 
and at a 2\% doping level ($n=0.49$) in (b) and (d).  
Parameter values used for calculation are: 
$V'/t=0.707$, $V/t=0.5$, and $t'/t=0.1$ in (a) and (b), and 
$V/t=1$ and $t'/t=0.1$ in (c) and (d).}
\end{center}
\label{fig:3}
\end{figure}
%%%%%%%%%%%%%%%%%%%%%%%%%%%%%%%%%%%%%%%%%%%%%%%%%%%%%%%%%%%%%%%%%%
Calculated results for the parameter and doping dependences 
of the optical conductivity spectra at $\omega>0$ for 
$E\parallel x$ are shown in Fig.~3.  We note that the spectral 
weight for $E\parallel y$ is two orders of magnitude smaller 
than that for $E\parallel x$ (and therefore not shown here) 
because the spectra for $E\parallel x$ have the weight roughly 
proportional to $t^2$ while the spectra for $E\parallel y$ have 
the weight proportional to $t'^2$, and here we assume the value 
$t'/t=0.1$.  
This is consistent with the fact that the experimental spectral 
weight for $E\parallel a$ is more than two orders of magnitude 
smaller than the weight for $E\parallel b$ \cite{takenaka}.  
Conversely, this agreement with experiment demonstrates that 
the value of $t'$ is indeed much smaller than the value of $t$ 
as we have assumed in this paper.  

We find in Fig.~3 that the spectra for the CO-2 ground state 
locate in the energy around $\omega/t=0.2-2$ whereas the 
spectra for the CO-1 ground state locate in a much higher 
energy range $\omega/t=3-6$.  Because the gaps of the 
quasiparticle band structure of the two CO states have 
a similar value, this difference in the location of the 
spectral weight is purely the matrix-element effects, i.e., 
the selection rule for these two types of CO states is 
quite different.  
We may have the following real-space interpretation for 
the obtained spectra: In the CO-1 ground state, the 
electric field along the chain takes out a hole and put it 
on the neighboring site to make either a doubly occupied 
site via $t$ or a singly occupied site via $t'$, the former 
of which costs the energy $U$ and the latter of which costs 
the energy $V'$.  Since we assume $t\gg t'$, the main 
spectral weight appears at the position $\omega\sim U$.  
We find in Figs.~3 (c) and (d) that the position of the 
main peak indeed depends strongly on $U$ but not on $V'$.  
In the CO-2 ground state, the excitation by the electric 
field along the chain does not create doubly occupied 
sites in the strong coupling limit, the energy cost 
of which is at most $V$.  We thus have the main peak 
much lower in energy.  
We find some fine structures of the spectra which 
reflects the fairly complicated quasiparticle multi-band 
structure.  

%%%%%%%%%%%%%%%%%%%%%%%%%%%%%%%%%%%%%%%%%%%%%%%%%%%%%%%%%%%%%%%%%%
\begin{figure}[t]
\vspace{10pt}
\begin{center}
\includegraphics[width=7.0cm,clip]{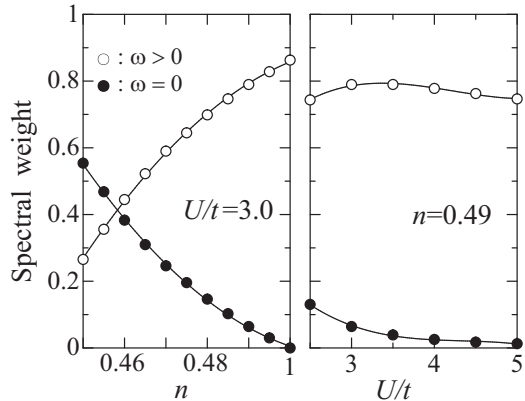}
\caption{Filling dependence (left panel) and $U/t$ 
dependence (right panel) of the spectral weight at 
$\omega=0$ (Drude weight, filled circles) and integrated 
weight over $\omega>0$ excluding the Drude contribution 
(open circles).  The CO-2 phase is assumed.  
Parameter values used for calculation are: 
$V/t=0.5$, $V'/t=0.707$, and $t'/t=0.1$.} 
\end{center}
\label{fig:4}
\end{figure}
%%%%%%%%%%%%%%%%%%%%%%%%%%%%%%%%%%%%%%%%%%%%%%%%%%%%%%%%%%%%%%%%%%%
Integrated spectral weight of the Drude ($\omega=0$) and 
interband ($\omega>0$) contributions are shown separately 
in Fig.~4.  We find that the Drude weight grows very 
rapidly by doping away from quarter filling at which 
the Drude weight vanishes.  We also find that the $U/t$ 
dependence of the integrated spectral weight is not very 
strong.  

\section{DISCUSSION}

Let us first summarize the experimental aspects of the 
optical conductivity of PrBa$_2$Cu$_4$O$_8$ observed at 
room temperature \cite{takenaka}.  The spectra have the 
following three features; 
(i) a broad peak structure with large spectral weight (98\% 
of the total weight) located around $\omega\simeq$40 meV, 
(ii) very small Drude weight (2\% of the total weight), and 
(iii) the spectra for $E\parallel a$ have the weight which 
is two orders of magnitude smaller the weight for 
$E\parallel b$.  
Another important aspect of the charge dynamics of this 
material is given by the NQR experiment \cite{fujiyama}; 
(iv) the CuO double chains have a very slow fluctuation of 
the electric-field gradient caused by the spatial 
fluctuation of electronic charge carriers.  The system may 
thus be in the vicinity of the long-range CO state.  

%%%%%%%%%%%%%%%%%%%%%%%%%%%%%%%%%%%%%%%%%%%%%%%%%%%%%%%%%%%%%%%%%%
\begin{figure}[t]
\vspace{10pt}
\begin{center}
\includegraphics[width=7.0cm,clip]{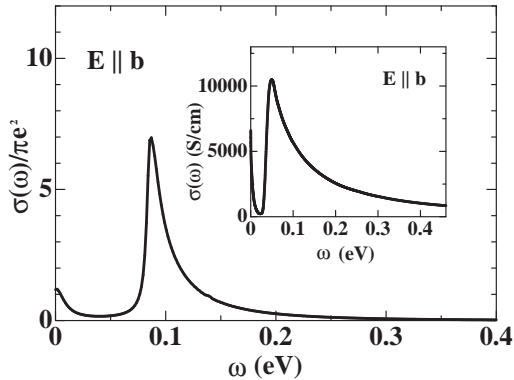}
\caption{Calculated optical conductivity spectrum compared 
with the experimental data of Takenaka 
{\it et al.} \cite{takenaka} (inset).  
Parameter values used for calculation are: 
$t=0.3$ eV, $U/t$=2.5, $V/t=0.5$, $V'/t=0.707$, $t'/t=0.1$, 
and $n=0.495$.  
The Drude peak and the peak at $\omega>0$ have the weight of 
6.7\% and 93.3\% of the total weight, respectively.  
The $\delta$-function of the Drude peak is broadened 
artificially with the Lorentzian of width $\eta/t=1/10\pi$.}
\end{center}
\label{fig:5}
\end{figure}
%%%%%%%%%%%%%%%%%%%%%%%%%%%%%%%%%%%%%%%%%%%%%%%%%%%%%%%%%%%%%%%%%%%
Then, in Fig.~5, we compare our calculated optical 
conductivity with the experimental spectrum 
\cite{takenaka} for PrBa$_2$Cu$_4$O$_8$.  
We find that the experimental spectrum is well reproduced; 
(i) The calculated spectrum, which is in reasonable agreement 
with experiment, is for the CO-2 ground state near the crossover 
region to the SDW+CDW phase (see Fig.~2).  
The spectra calculated for the CO-1 states completely fail 
to agree with experiment as already pointed out in Sec.~IV.  
(ii) The spectra for the electric field perpendicular 
to the chains have the weight two orders of magnitude smaller 
than those for the electric field parallel to the chains, 
which is consistent with experiment (see Sec.~IV).  
(iii) The calculated peak structure is found to locate at 
rather higher energies than experiment.  Although this is 
partly due to insufficiency of the mean-field 
approximation, fine-tuning of the parameter values would 
improve the agreement with experiment.  
(iv) The observed very small Drude weight is well reproduced 
by our calculation.  
(v) The asymmetric spectral shape of the main peak structure, 
i.e., sharp in the lower energy side and broad in the higher 
energy side, is well reproduced by our calculation.  
The width of the peak is however rather narrow compared with 
experiment.  This may be due to the effect of electron 
correlations neglected in the mean-field approximation, 
i.e., the scattering of the quasiparticles may well broaden 
the spectra \cite{takenaka}.  
(vi) As for the amplitude of the ordered charges, we find 
the calculated values of the order parameters to be 
$\delta=0.013$ (per site) and $2S^z=0.21$ (per site) for 
the parameter values used in the calculation of the optical 
conductivity of Fig.~5.  
This value of $\delta$ is consistent with the experimental 
estimation where it has been pointed out that less than 
0.02 holes per Cu are frozen \cite{fujiyama}.  

Finally, let us discuss the optical conductivity spectra 
for fluctuating charge carriers.  We should first note 
that there is no experimental indication of the long-range 
CO in PrBa$_2$Cu$_4$O$_8$ but rather there is only the slow 
fluctuation of electronic charge degrees of freedom 
\cite{takenaka}.  Then, in this situation, we find that the 
spectra calculated for the {\it ordered} state agree 
reasonably well with experimental spectra.  
The time-scale argument may readily resolve this apparent 
contradiction; the fast time scale of the optical measurement 
should enable one to detect slowly fluctuating order 
parameters as virtually a long-range order.  
This situation reminds us of the optical conductivity spectra 
of the well-known CO material $\alpha'$-NaV$_2$O$_5$ 
\cite{ohama}.  The spectra for the CO phase of this system 
differ very little from the spectra of the uniform phase at 
temperatures above $T_{\rm CO}$; there exists slow fluctuation 
of the charge degrees of freedom even far above the transition 
temperature \cite{nishimoto1,nishimoto2}.  
Similar situation has also been noticed for the quasi-1D 
organic materials \cite{tajima}.  

In the present CuO double-chain systems, if the system were 
exactly at quarter filling, the long-range CO might occur 
with vanishing metallic conductivity.  
This means that if a fine filling-control is made on the CuO 
double chains one may be able to obtain the insulating double 
chains.  We therefore would suggest that the experimental 
filling control, e.g., by applying pressure \cite{yamada} or 
elemental substitution \cite{nakada} (if it purely changes 
the filling of the CuO double chains), seems to be quite 
interesting.  

\section{SUMMARY}

We have considered the anomalous charge dynamics of the 
CuO double chains of PrBa$_2$Cu$_4$O$_8$ by focusing on 
the optical conductivity spectra recently observed in 
experiment \cite{takenaka}.  

Our study presented in this paper may be summarized 
as follows: 
(i) We have experimental evidence of the presence of strong 
charge fluctuations, i.e., the system may be in the vicinity 
of the charge-ordered state.  
Then, one may naturally assume the presence of strong 
intersite Coulomb repulsions in the system.  
(ii) We have therefore adopted an extended Hubbard-type model 
with the intersite Coulomb repulsion near quarter filling 
and treated this in the mean-field approximation.  We thus 
have found that the possible charge-ordered state may be either 
of the in-line-type or of the stripe-type charge modulations.  
(iii) We have calculated the optical conductivity spectra for 
the model in the same approximation and found that the 
spectra calculated for the stripe-type charge ordering 
reproduce the experimental features of the $\sim$40 meV 
peak and small Drude weight reasonably well but that the 
spectra for the in-line-type charge ordering does not agree 
with experiment.  
(iv) Then, we have argued that the stripe-type fluctuations of 
charge carriers should present in the system and this is the 
origin of the anomalous charge response of the system.  
The CuO double chains have no {\it long-range} charge ordering 
but the fast time-scale of the optical measurement should 
detect the very slow fluctuation of charges as virtually 
a long-range charge ordered state.  

The CuO double chains in PrBa$_2$Cu$_4$O$_8$ thus provide 
a good opportunity for studying the anomalous charge 
dynamics in the 1D strongly correlated electron systems.  
We hope that further experimental and theoretical studies 
will be made to clarify various aspects of this intriguing 
system.  

\begin{acknowledgments}
We thank K. Takenaka for sending us their unpublished 
experimental data and enlightening discussion.  We also 
acknowledge useful discussions with S. Horii, T. Ohama, 
H. Seo, and H. Yoshioka.  
This work was supported in part by Grants-in-Aid for 
Scientific Research (Nos.~11640335 and 12046216) from the 
Ministry of Education, Culture, Sports, Science, and 
Technology of Japan.  
Computations were carried out at the computer centers of 
the Institute for Molecular Science, Okazaki, and the 
Institute for Solid State Physics, University of Tokyo.  
\end{acknowledgments}


\begin{thebibliography}{9}

\bibitem{horii1} S. Horii, U. Mizutani, H. Ikuta, Y. Yamada, 
J. H. Ye, A. Matsushita, N. E. Hussey, H. Takagi, and 
I. Hirabayashi, Phys. Rev. B {\bf 61}, 6327 (2000).

%Y-Pr1248: synthesis. 
\bibitem{horii2} S. Horii, Y. Yamada, H. Ikuta, N. Yamada, 
K. Kodama, S. Katano, Y. Funahashi, S. Morii, A. Matsushita, 
T. Matsumoto, I. Hirabayashi, and U. Mizutani, 
Physica C {\bf 302}, 10 (1998).  

\bibitem{fehrenbacher} R. Fehrenbacher and T. M. Rice, 
Phys. Rev. Lett.{\bf 70}, 3471 (1993).

%Pr1248 optical conductivity. 
\bibitem{takenaka} K. Takenaka, K. Nakada, A. Osuka, 
S. Horii, H. Ikuta, I. Hirabayashi, S. Sugai, and 
U. Mizutani, Phys. Rev. Lett. {\bf 85}, 5428 (2000).  

%Pr1248ARPES
\bibitem{mizokawa} T. Mizokawa, A. Ino, T. Yoshida, 
A. Fujimori, C. Kim, H. Eisaki, Z.-X. Shen, S. Horii, 
T. Takeshita, S. Uchida, K. Tomimoto, S. Tajima, and 
Y. Yamada, {\it Stripes and Related Phenomena} (Selected 
Topics in Superconductivity, Vol. 8) edited by A. 
Bianconi and N. L. Saini, (Plenum, New York, 2000). 

%Pr1248: NMR
\bibitem{fujiyama} S. Fujiyama, M. Takigawa, and 
S. Horii, unpublished.  

%Optical conductivity 
\bibitem{tajima} H. Tajima, Solid State Commun. {\bf 113}, 
279 (2000). 

%TMTTF paper
\bibitem{shibata} Y. Shibata, S. Nishimoto, and Y. Ohta, 
Phys. Rev. B {\bf 64}, to appear in 15 Nov. issue (2001).  

%Pr1248: theory
\bibitem{seo} H. Seo and M. Ogata, Phys. Rev. B {\bf 64}, 
113103 (2001). 

%basic theories
\bibitem{psaltakis} G. C. Psaltakis, J. Phys.: Condens. 
Matter {\bf 8}, 5089 (1996).  
\bibitem{bari} R. A. Bari, D. Adler, and R. V. Lange, 
Phys. Rev. B {\bf 2}, 2898 (1970). 

\bibitem{ohama} T. Ohama, H. Yasuoka, M. Isobe, and 
Y. Ueda, Phys. Rev. B {\bf 59} 3299 (1999).  

\bibitem{nishimoto1} S. Nishimoto and Y. Ohta, 
J. Phys. Soc. Jpn. {\bf 67}, 3679 (1998).

\bibitem{nishimoto2} S. Nishimoto and Y. Ohta, 
J. Phys. Soc. Jpn. {\bf 67}, 4010 (1998).

%Pressure-induced charge flow in Y1248. 
\bibitem{yamada} Y. Yamada, J. D. Jorgensen, S. Y. Pei, 
P. Lightfoot, Y. Kodama, T. Matsumoto, and F. Izumi, 
Physica C {\bf 173}, 185 (1991).  

%Pr1248 Zn doping. 
\bibitem{nakada} K. Nakada, H. Ikuta, S. Horii, 
H. Hozaki, I. Hirabayashi, and U. Mizutani, 
Physica C {\bf 357-360}, 186 (2001). 

\end{thebibliography}
\end{document}